\documentclass[conference]{IEEEtran}
\IEEEoverridecommandlockouts
\usepackage{cite}
\usepackage{amsmath,amssymb,amsfonts}
\usepackage{algorithmic}
\usepackage{graphicx}
\usepackage{textcomp}
\usepackage{xcolor}
\usepackage{amsmath}
\usepackage{fullpage}
\usepackage{times}
\usepackage{fancyhdr,graphicx,amsmath,amssymb}
\usepackage[ruled,vlined]{algorithm2e}       
\usepackage{subcaption}
\usepackage{mathrsfs}

\def\BibTeX{{\rm B\kern-.05em{\sc i\kern-.025em b}\kern-.08em
    T\kern-.1667em\lower.7ex\hbox{E}\kern-.125emX}}
\begin{document}

\title{Augmenting SEFDM Performance in High-Doppler Channels\\
}

\author{\IEEEauthorblockN{1\textsuperscript{st} Mahdi Shamsi}
	\IEEEauthorblockA{\textit{EE. dept. of Sharif Uni. of Tech.} \\
		\textit{Advanced Comm. Research Institute (ACRI).}\\
		\textit{Multimedia and Signal processing Lab. (MSL).}\\
		Mahdi.Shamsi@alum.sharif.edu}
	\and
	\IEEEauthorblockN{2\textsuperscript{nd} Farokh Marvasti}
	\IEEEauthorblockA{\textit{EE. dept. of Sharif Uni. of Tech.} \\
		\textit{Advanced Comm. Research Institute (ACRI).}\\
		\textit{Multimedia and Signal processing Lab. (MSL).}\\
		marvasti@sharif.edu}
	}
\maketitle

\begin{abstract}
This paper presents an innovative approach leveraging Spectrally Efficient Frequency Division Multiplexing (SEFDM) with enhancements, including Frequency Domain Cyclic Prefix (FDCP) and Modified Non-Linear (MNL) acceleration, to address challenges arising from delay and Doppler shift in mobile communication channels. Our methodology demonstrates superior performance and enhanced spectral efficiency while incurring minimal computational overhead. We conduct a rigorous evaluation of the SEFDM system model, examining the impact of the MNL acceleration on SEFDM detection, and assessing the effectiveness of FDCP in mitigating Doppler shift effects. Bit Error Rate (BER) results obtained through Regularized Sphere Decoding in diverse simulation scenarios affirm the capability of our proposed solution to effectively manage challenges, ensuring robust and high-quality connectivity in mobile communication networks.
\end{abstract}

\begin{IEEEkeywords}
Overloaded modulation, SEFDM, High-Doppler Channels, Inverse systems, Spectral Efficiency.
\end{IEEEkeywords}

\section{Introduction}
In response to the growing emphasis on enhancing spectral efficiency, Spectrally Efficient Frequency Division Multiplexing (SEFDM) has been introduced, drawing inspiration from the Mazo limit concept \cite{mazo1975faster} as outlined in \cite{rodrigues2003spectrally}. The deployment of SEFDM aims to retain the favorable attributes of digital Orthogonal Frequency Division Multiplexing (OFDM), facilitating the seamless transition of OFDM-based systems to SEFDM with minimal hardware costs.

In contemporary communication environments characterized by high population density and increased mobility, the reception of signals by the receiver may incur substantial time delay and Doppler shift effects, leading to performance degradation and suboptimal user experiences. This phenomenon is particularly pronounced in mobile sensor networks, where such distorted communication has the potential to propagate errors, significantly impacting overall network performance \cite{shamsi2021flexible,shamsi2023distributed}. Addressing this challenge, Orthogonal Time-Frequency Space (OTFS), as introduced in \cite{rodrigues2003spectrally}, employs a Delay-Doppler (DD) representation of both the signal and the channel. However, the two-dimensional nature of the OTFS introduces higher delays, and fractional shifts in the DD domain necessitate additional considerations, as highlighted in \cite{lampel2021orthogonal}. To surmount these limitations and effectively manage the effects of delay and Doppler shift, we propose the utilization of SEFDM, complemented by advanced techniques such as Frequency Domain Cyclic Prefix (FDCP) \cite{dean2020rethinking} and Modified Non-Linear (MNL) acceleration \cite{shamsi2020nonlinear,shamsi2024acceleration}.

Our technical methodology demonstrates superior performance and spectral efficiency when compared to conventional communication systems, all while maintaining low computational costs. Our focus is on developing highly efficient solutions to address the challenges posed by delay and Doppler shift in mobile communication systems, ensuring reliable and high-quality communication even in extremely challenging environments.
In summary, our key contributions encompass:

\begin{itemize}
	\item Strengthening the robustness of SEFDM against high-Doppler channels through the implementation of FDCP-based SEFDM.
	\item Enhancing the performance of SEFDM in terms of convergence and stability of the inverse system-based SEFDM.
	\item Streamlining the complexity associated with fractional shifts in the DD domain.
\end{itemize}

The current manuscript is organized in the following manner. Section \ref{sect:Model} introduces SEFDM model along with proposing the use of cyclic prefix (CP) in the frequency domain and describing the detection algorithms utilized. Detailed reporting of the simulation results is provided in Section \ref{sect:sim}. Finally, the paper is concluded in Section \ref{sect:con}.
\section{SEFDM Model}
\label{sect:Model}
\begin{figure*}[h]
	{\includegraphics[width=\textwidth,height=.5\textwidth]{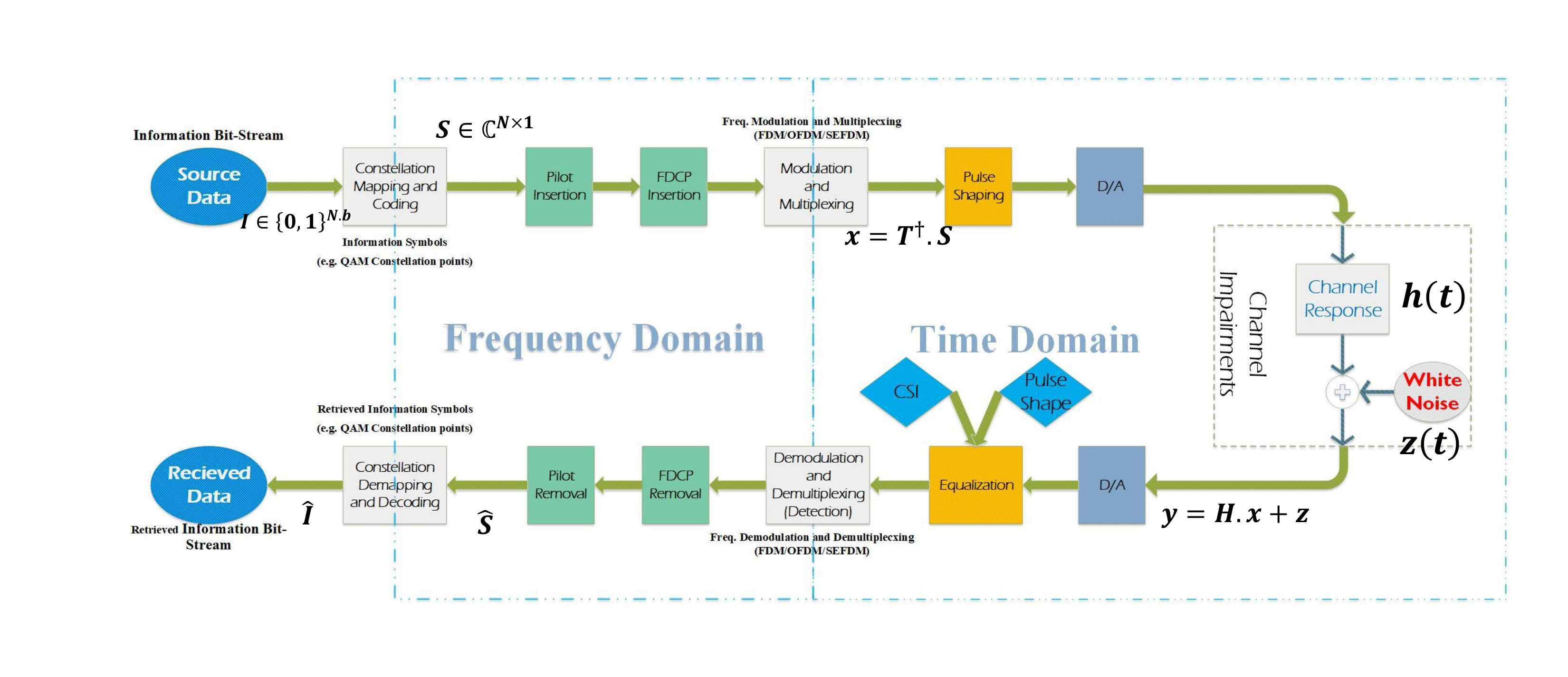}}
	\caption{Block-diagram of telecommunication system {MCM} based on frequency division with {FDCP} and pulse shaping.}
	\label{blk_FDCP}
\end{figure*}

By specifying the compression factor $\alpha \in (0, 1]$ as a parameter, the Spectrally-Efficient Frequency Division Multiplexing (SEFDM) technique can be formulated as the following equation:
\begin{equation}
	x[n] = \sum_{k=0}^{N-1} S[k] \, e^{j 2 \pi \alpha\frac{n k}{N}},
	\label{eq:SEFDMDef}
\end{equation}
where $x[n]$ is the sampled signal in the time domain, while $S[k]$ are the frequency-domain complex symbols. SEFDM's process can be illustrated using the block diagram shown in Fig. \ref{blk_FDCP}, which represents the baseband of frequency-based telecommunication systems. The input data stream ($\boldsymbol{I}$) with $b$ bits per symbol (in this example, $b=2$ for 4-QAM) is mapped to $N.b$ constellation points in a frame interval using $\boldsymbol{S}$. This vector is then transformed from the frequency domain to the time domain using an $N\times N$ sized SEFDM matrix, designated as $T$, which follows \eqref{eq:SEFDMDef}.

After the signal is transmitted through the communication channel, we encounter delayed versions of the signal, each of which may have experienced a different Doppler shift. The transmitted signal $x(t)$ is received as:
$y(t) = \sum_{i=1}^{P} h_i e^{2\pi j \nu_i (t-\tau_i)} x(t-\tau_i) + z(t)$,
where $z(t)$ represents a white Gaussian noise and $h_i$, $\tau_i$, and $\nu_i$ denote the attenuation, delay, and Doppler shift of the $i$-th path, according to the definition of the channel response.
After performing the detection algorithm, an estimation of the input symbol signal ($\hat{\boldsymbol{S}}$) and thus an estimation of the binary input information flow ($\hat{\boldsymbol{I}}$) is obtained.
\subsection{Adding Cyclic Prefix in the Frequency Domain}
Multi-Carrier Modulation (MCM) enables compensation for the channel response comprising diverse delayed paths. By incorporating a Cyclic Prefix (CP) into the transmitted time frame, DFT-based analysis is rendered feasible, and Inter-symbol Interference (ISI) caused by multipath interference takes the form of a single coefficient in the frequency domain \cite{goldsmith2005wireless,tse2005fundamentals}. It is essential to bear in mind that, in this technique, the CP length must exceed the longest channel impulse response in the time domain. Motivated by this approach, we adopted a similar dual strategy to that presented by the authors in \cite{dean2020rethinking} in which we introduced a CP in the frequency domain to counteract the adverse impact of receiving multiple signal versions in a Doppler channel.

In other words, based on the input-output relationship of the channel and by defining an intermediary variable $u(t)$:
\begin{eqnarray}
\{u_i(t)\triangleq e^{2\pi i \nu_i t}x(t);\;X(f)= \mathscr{F}\{x(t)\}\notag\\
\Rightarrow	U_i(f) = \mathscr{F} \{u_i(t)\} = X(f-\nu_i),	\\
	\begin{Bmatrix}
		y(t)=\sum_{i=1}^{P} h_i u(t-\tau_i)\\ 
		\mathscr{F} \{u_i(t-\tau_i)\}= e^{-2\pi j f\tau_i}U_i(f)
	\end{Bmatrix}\notag\\
\Rightarrow  Y(f) = \sum_{i=1}^{P} h_i e^{-2\pi j f\tau_i}U_i(f),
\end{eqnarray}
we can describe the above cases as follows. Consider a Doppler-free channel ($\nu_i=0$ and $u_i(t)=x(t)$) with suitable CP, such that DFT-based analysis is valid. In this case, with a discrete signal representation, we have: $Y[k] = \sum_{i=1}^{P} h_i e^{-2\pi j k\tilde{\tau}_i}X[k];\;\; \tilde{\tau}_i = \frac{\tau_i}{T_s}.$

We can see that each frequency component $X[k]$ experiences the channel as a single coefficient $H[k]\triangleq\sum_{i=1}^{P} h_i e^{-2\pi j k\tilde{\tau}_i}$. Compensation for such a channel, with having perfect CSI, is achieved by simply dividing component by component: $R[k]\triangleq X[k]/H[k]$.\footnote{It should be noted that adding a CP to the transmitted signal has converted the linear convolutions to circular convolutions and enabled the use of DFT.}

We can consider the Doppler channel in a dual form. In such a channel, different versions of the transmitted signal with different Doppler shifts are received without delay spread ($\tau_i=0$ and $Y(f) = \sum_{i=1}^{P} h_i X_i(f-\nu_i)$). In this case, after frequency shifting the transmitted signal equivalent to $L$ components after sampling, it is added to the end of the signal equal to that amount:
\begin{align}
	&\tilde{X}(f)&\triangleq&\left\{\begin{matrix}
		&X(f-L.\Delta f)&;&\; L.\Delta f\leq f<BW\\ 
		&X(f+BW-L.\Delta f)&;&\; 0\leq f<L.\Delta f
	\end{matrix}\right.\notag\\
	&\tilde{Y}(f)& =& \sum_{i=1}^{P} h_i \tilde{X}_i(f-\nu_i)\notag\\
	&\Rightarrow& &Y(f)=\left\{\begin{matrix}
		\tilde{Y}(f+L.\Delta f);\; 0\leq f<BW-L.\Delta f\\ 
		0;\; else
	\end{matrix}.\right. \notag
\end{align}
Thus, if $\nu_P=\max_i(\nu_i)<L.\Delta f$ is satisfied, by taking $L = BW/\Delta f$, we can write the received signal in the receiver as: $Y[k]=X[k]\circledast H[k];\;k = 0,1,\dots,N-L-1,$ where $\circledast$ is a circular convolution. In this case, interference from other frames is removed from the current frame. Fig. \ref{blk_FDCP} shows a block diagram of the frequency-division multiplexing system with the addition of a frequency-domain CP and pulse shaping for better intuition.

\subsection{Detection: Sphere Decoding and Inverse Systems}
We report the performance of SEFDM detection algorithms in some scenarios using both Iterative Method (IM) detection and regularized sphere decoding (RegSD) and report their BER. First, we consider the IM algorithm as the SEFDM detection algorithm and use both hard and soft decisions for calculating an estimated transmitted symbol. We perform 4-QAM modulation and calculate the output component $s_i$ of the $i$-th symbol. To estimate the signal in the $k$-th iteration ($\boldsymbol{u}^{<k>}$) using the recursive description of the IM, with initial input $\boldsymbol{u}^{<0>} = T\boldsymbol{y}$ and a total of $\eta$ iterations, we have \cite{xu2013improved}:
$\boldsymbol{u}^{<k>} \leftarrow \lambda(\boldsymbol{u}^{<0>} -G^{\text{{soft}}}(\boldsymbol{u}^{<k-1>},d))+\boldsymbol{u}^{<k-1>}$, $d \leftarrow 1-k/\eta.$
It is clear that with placing $d$ at the decision boundary of the transmitted symbols, the corresponding channel decoder operator can be defined as hard decision ($ G^{\text{{hard}}}(\boldsymbol{u})\triangleq G^{\text{{soft}}}(\boldsymbol{u},d\leftarrow0)$).

To enhance detection performance, we adopt the MNL method presented in \cite{shamsi2020nonlinear} and the RegSD MATLAB built-in function that employs single-tree search according to the algorithm proposed in \cite{studer2008soft}. The MNL method accelerates convergence by utilizing four successive estimated symbols, $\hat{x}_0, \dots \hat{x}_3$:
\begin{equation}
	x_{_{MNL}}=\begin{cases}
		\frac{\hat{x}_3\times\hat{x}_1-\hat{x}_2^2}{\hat{x}_3+\hat{x}_1-2\hat{x}_2}\:;\: \vert\sigma_0\vert\leq\vert\sigma_1\vert\\
		
		\frac{\hat{x}_2\times\hat{x}_0-\hat{x}_1^2}{\hat{x}_2+\hat{x}_0-2\hat{x}_1}\:;\: \vert\sigma_0\vert>\vert\sigma_1\vert
	\end{cases},
\end{equation}
where $\hat{x}_i$ is an arbitrary estimated symbols in the $i^{\text{th}}$ iteration and $\sigma_1\triangleq\hat{x}_3+\hat{x}_1-2\hat{x}_2\,,\,\sigma_0\triangleq\hat{x}_2+\hat{x}_0-2\hat{x}_1$.
\section{Simulation Results}
\label{sect:sim}
In this section, we evaluate the performance of SEFDM detection algorithms in AWGN and delay-Doppler channels by measuring Bit Error Rate (BER) as a function of EB/N0 through simulating various scenarios. For a delay-Doppler channel simulation, we use  a Tapped Delay Line model (TDL) and refer to Table 7.7.2-1 in \cite{3gppChannel} for different delay scales proposed, which cover short to long delay spreads.

\begin{figure}[!hbt]
	\centering
	\begin{subfigure}[t]{.5\textwidth}
		\includegraphics[trim={0 0 0 0}, scale=0.275]{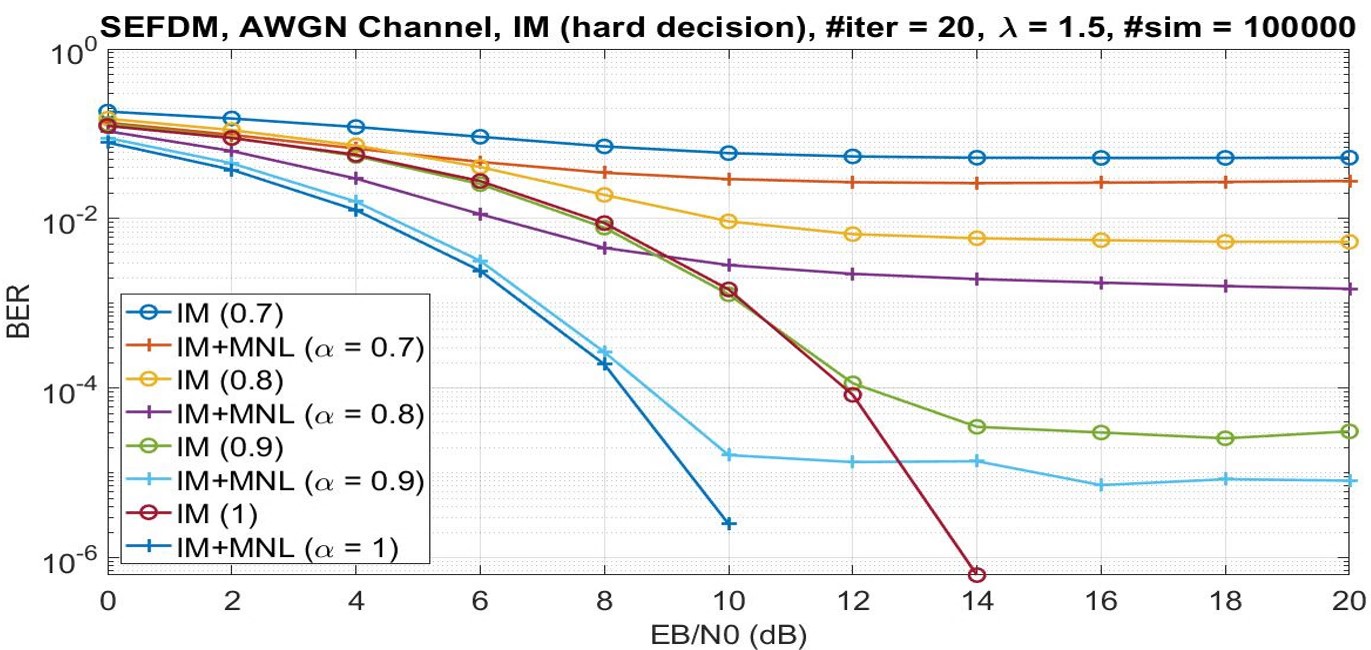}
		\caption{Hard decision.}
		\label{fig:SEFDM_AWGN_11}
	\end{subfigure}%
	\hfill
	\begin{subfigure}[t]{.5\textwidth}
		\includegraphics[trim={0 0 0 0}, scale=0.275]{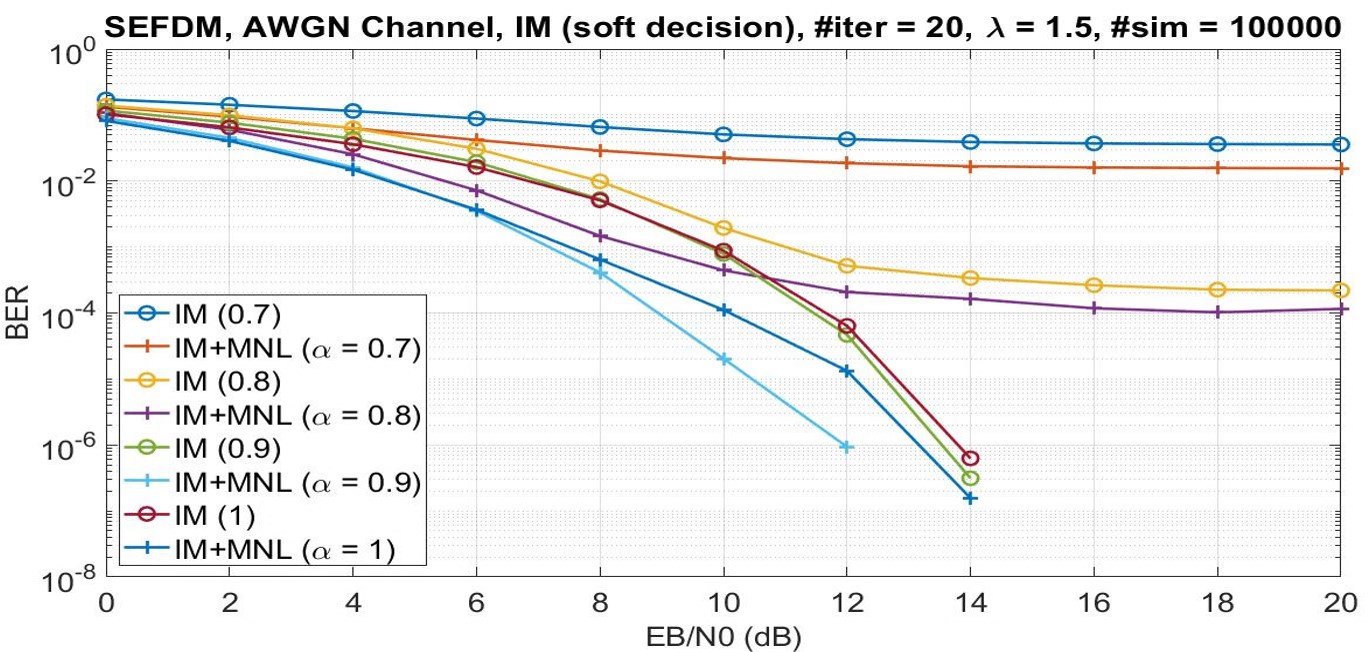}
		\caption{Soft decision.}
		\label{fig:SEFDM_AWGN_12}
	\end{subfigure}%
	\caption{
		Signal detection
		{SEFDM}
		With the help of algorithm
		{IM+MNL} in the AWGN channel.}
	\label{fig:SEFDM_AWGN}
\end{figure}

Figure \ref{fig:SEFDM_AWGN} demonstrates the effectiveness of using the MNL method to accelerate the IM in SEFDM under AWGN conditions. Both hard and soft decision IM can be improved with the help of the MNL method. Using the MNL method can also reduce the sensitivity of the IM to changes in the relaxation parameter ($\lambda$) and extend its stability range. 

\begin{figure}[!hbt]
	\centering
	\begin{subfigure}[t]{.5\textwidth}
		\includegraphics[trim={0 0 0 0}, scale=0.17]{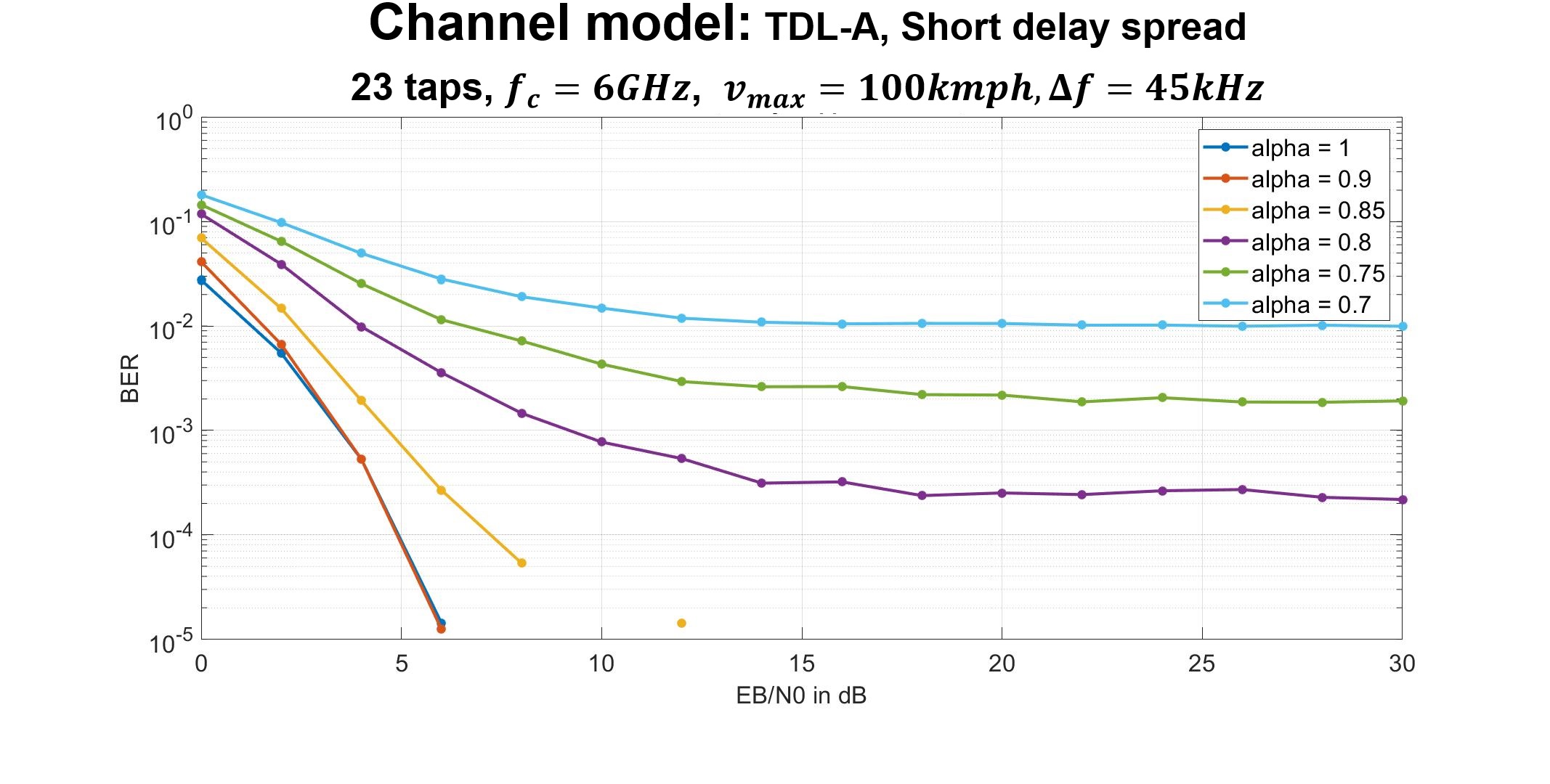}
		\caption{
			short delay spread.}
		\label{fig:SEFDM_DD_DS_11}
	\end{subfigure}%
	\hfill
	\begin{subfigure}[t]{.5\textwidth}
		\includegraphics[trim={0 0 0 0}, scale=0.17]{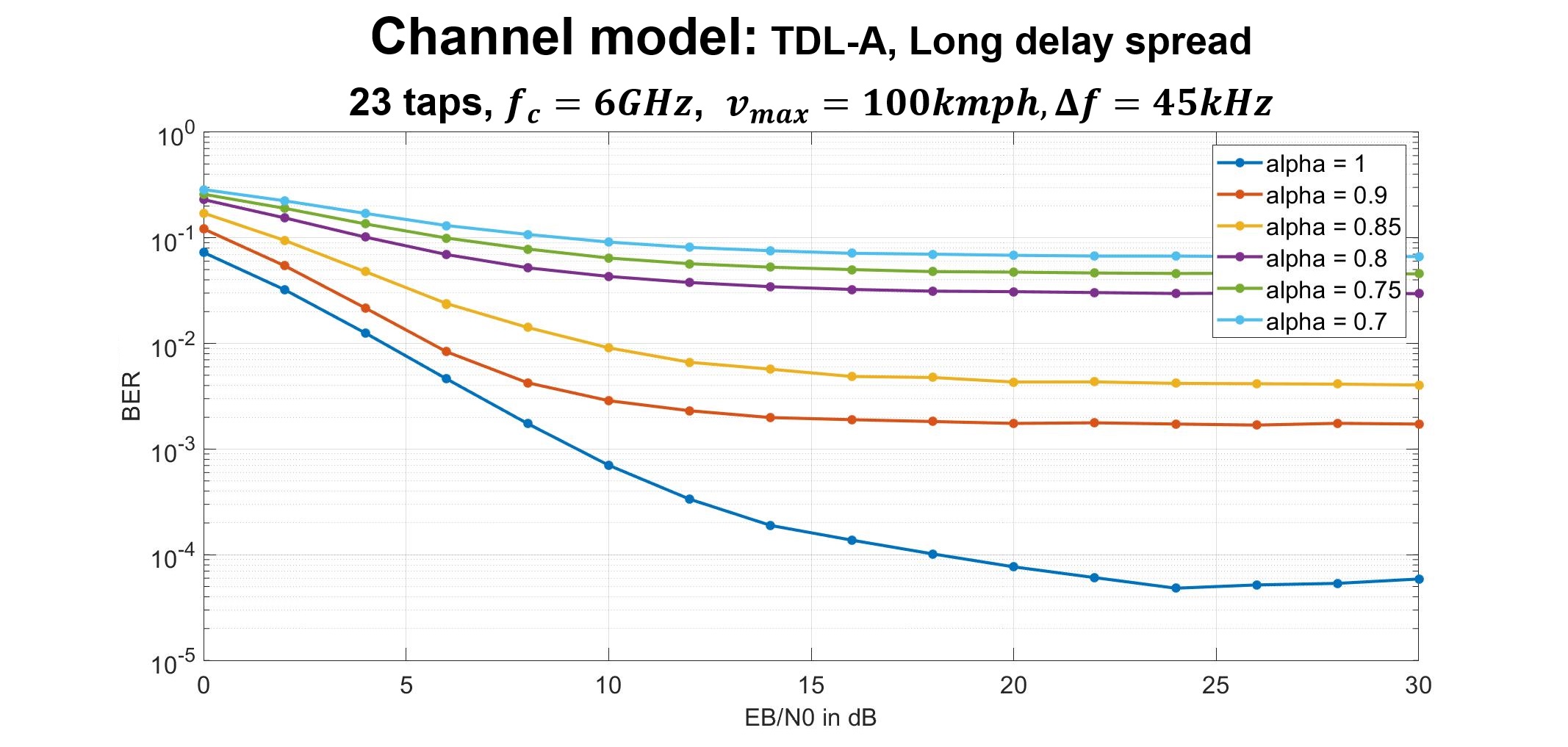}
		\caption{
			Long delay spread.}
		\label{fig:SEFDM_DD_DS_12}
	\end{subfigure}%
	\caption{
		Signal detection
		{SEFDM}
		With the help of algorithm
		{IM}
		with soft decision making in the delay-Doppler channel:
		{TDL-A} model with $v_{max}=100kmph$ and 23 delay circuits,
		$f_c=6GHz$, $T\approx 347ns$ $ (\Delta f = 45 kHz)$.
		\newline
		Each transmitted {SEFDM} frame:
		$N=64$, $CP = 8$, ideal {CSI},
		the system
		{4-QAM},
		Pulse shaping with
		{RRC} ($\beta = 0.25, sps = 10, span = 16$),
		The number of simulation iterations $=$10000.
		\newline
		Algorithm parameters
		{IM}:
		$\lambda = 0.8$ and $n_{iter} = 50$.
	}
	\label{fig:SEFDM_DD_DS}
\end{figure}

In Fig. \ref{fig:SEFDM_DD_DS}, we considered the entire 23-tap channel delay model and repeated the simulation for both short delay spread (Fig. \ref{fig:SEFDM_DD_DS_11}) and long delay spread (Fig. \ref{fig:SEFDM_DD_DS_12}).
As expected, in the case of short delay spread, the IM detector successfully recovered the transmitted symbols with negligible error ($BER\approx2\times10^{-5}$) at $EB/N0=6dB$ for $\alpha=0.9$.
In fact, the SEFDM performance in this case was similar to the OFDM.
However, in the case of long delay spread, the BER curves reached saturation.
It can be concluded that the examined idea (adding FDCP) compensated for the effect of Doppler shift but an increase in delay spread can compromise system performance.

In order to improve the detection performance, an SD-based approach was employed. The TDL-A model has been used in the simulations in a carrier frequency of $6 GHz$. The length of each SEFDM frame is considered to be $16$, with the data symbols and FDCP components constituting $3$ and $13$ components, respectively. The distance between each subcarrier is considered to be $\Delta f = 45 kHz$, which means that the frame length is approximately $1.4 \mu s$. To shape the pulse, RRC with $\beta = 0.25$, $sps = 10$, and $span = 16$ has been used. Results are reported by averaging over $10000$ different realizations of data signals, noise, and channel response.

\begin{figure*}[hbt]
		\begin{subfigure}[hbt]{\textwidth}
		\centering
		\includegraphics[trim={0 0 0 0}, scale=0.3]{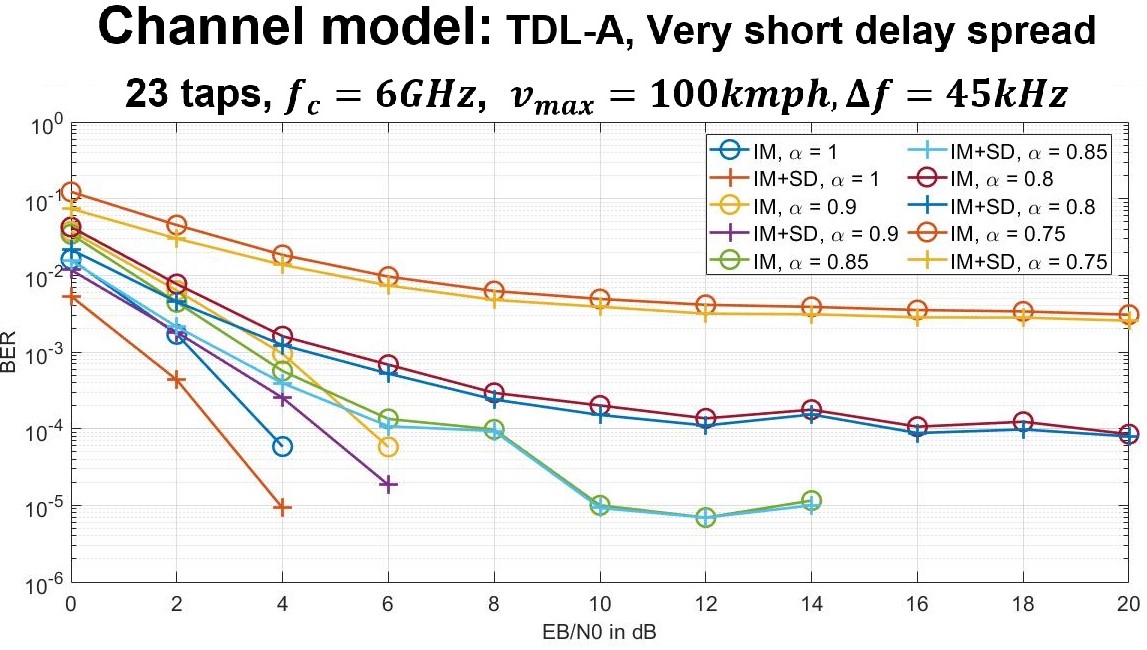}
		\caption{
			Soft decision making, $v_{max}= 100kmph$.
		}
		\label{fig:SEFDM_DD_SD2}
	\end{subfigure}
	\begin{subfigure}[hbt]{\textwidth}
		\centering
		\includegraphics[trim={0 0 0 0}, scale=0.2]{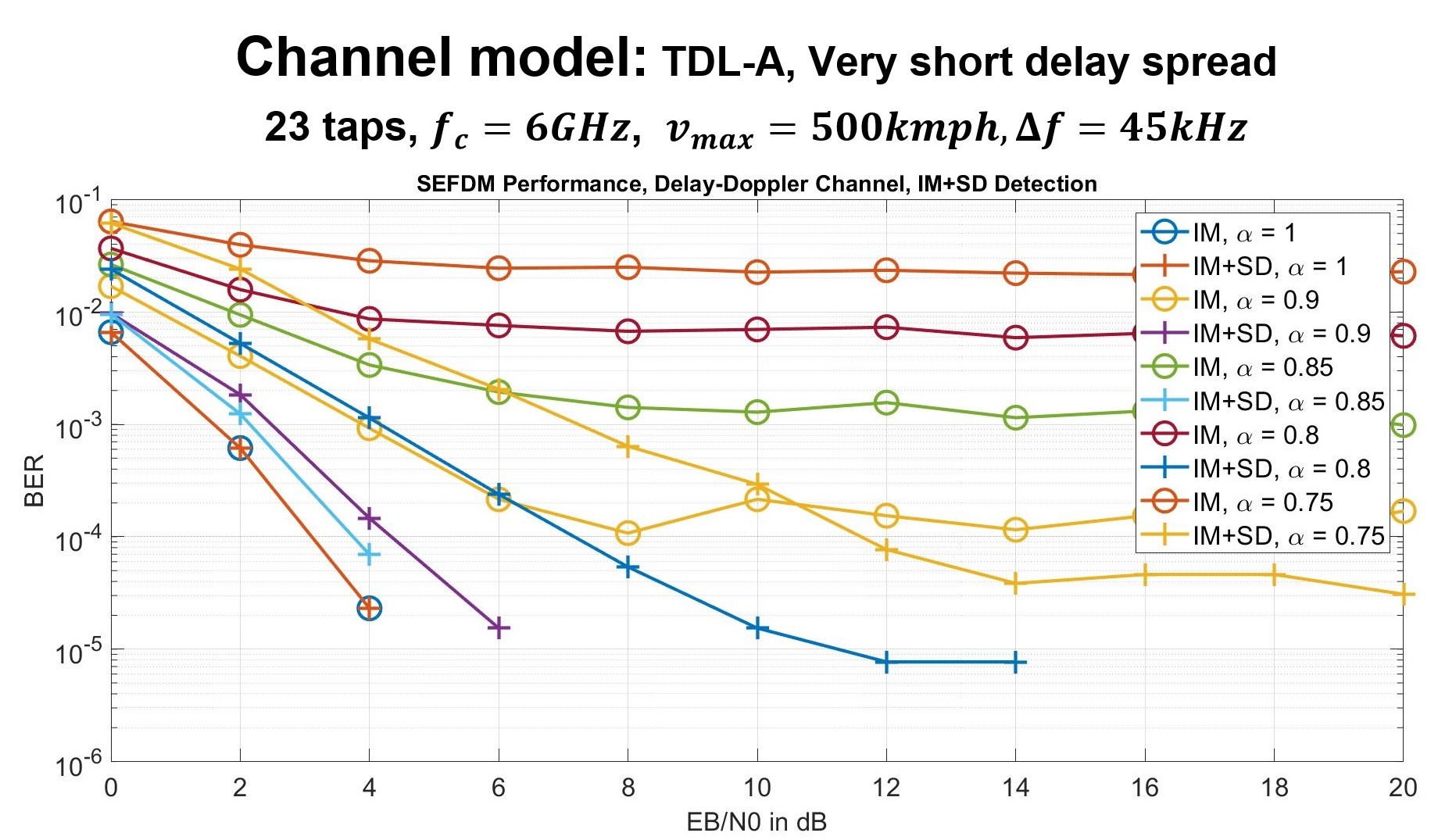}
		\caption{
			Hoft decision making, $v_{max}= 500kmph$.
		}
		\label{fig:SEFDM_DD_SD9}
	\end{subfigure}

\caption{
	Plot of {BER} in terms of {EB/N0} in signal detection
	{SEFDM}
	Using {IM} and {SD} algorithm in delay-Doppler channel:	$\lambda = 1$ and $n_{iter} = 20$.
}
	\end{figure*}

In Fig. \ref{fig:SEFDM_DD_SD2}, soft decision alleviates the impact of short delay spread at a maximum velocity of 100 km/h. However, increased iterations of the IM may compromise the SD efficacy, particularly with $\alpha=0.8,0.75$. Fig. \ref{fig:SEFDM_DD_SD9} illustrates the efficacy of SD with the hard decision IM under very short delay spread and high Doppler shift ($v_{max}=500 \, \text{km/h}$). Here, the SD importance is evident, and the IM iterations are minimized to prevent SD overload, resulting in significantly reduced BER across various SEFDM compression values.

\begin{figure*}[hbt]
	\begin{subfigure}{\textwidth}
		\centering
		\includegraphics[trim={0 0 0 0}, scale=0.2]{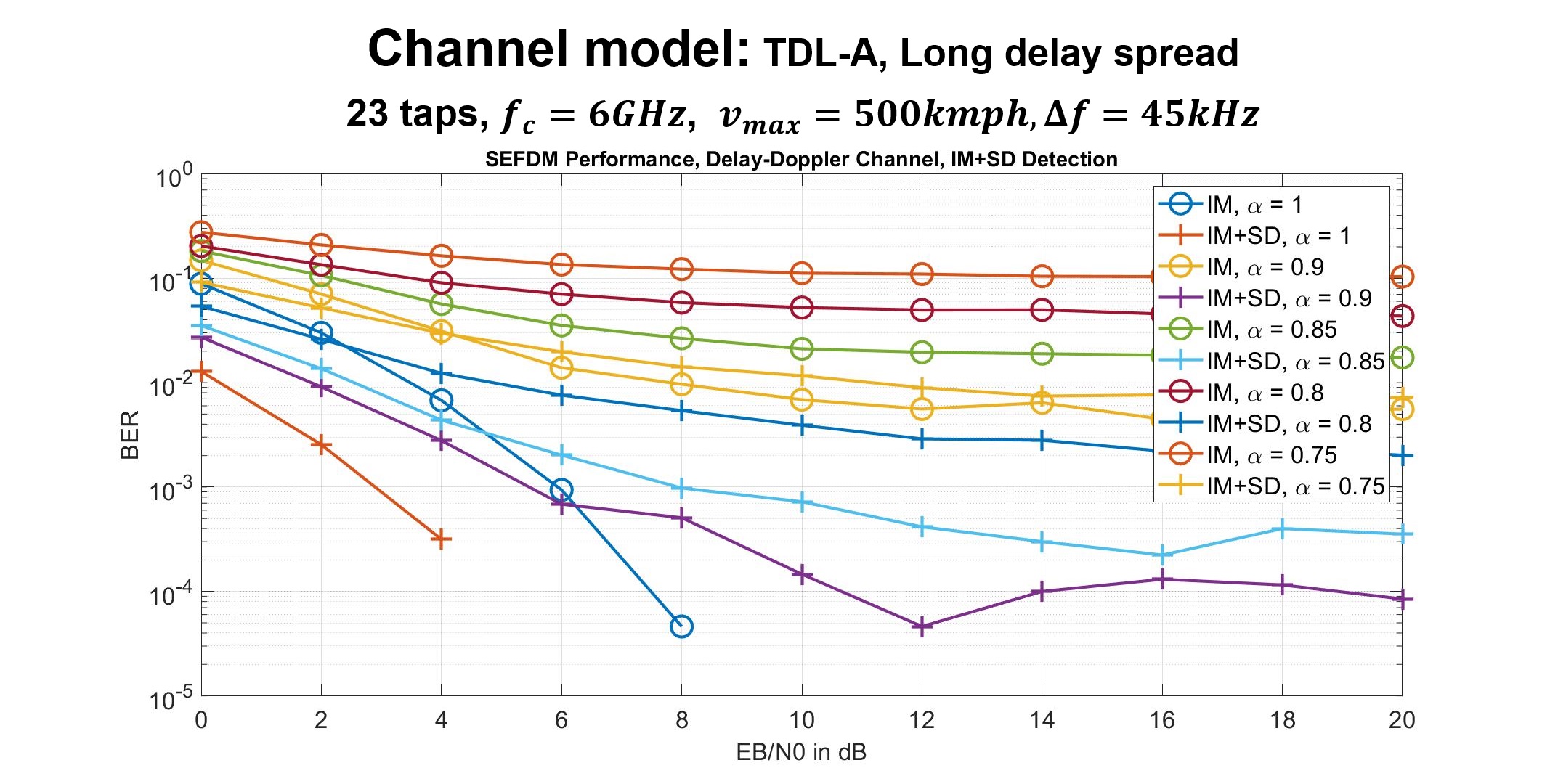}
		\caption{Soft decision making with
			$\lambda = 1$ and $n_{iter} = 5$.
		}
		\label{fig:SEFDM_DD_SD3}
\end{subfigure}%
\hfill
\begin{subfigure}{\textwidth}
	\centering
		\includegraphics[trim={0 0 0 0}, scale=0.2]{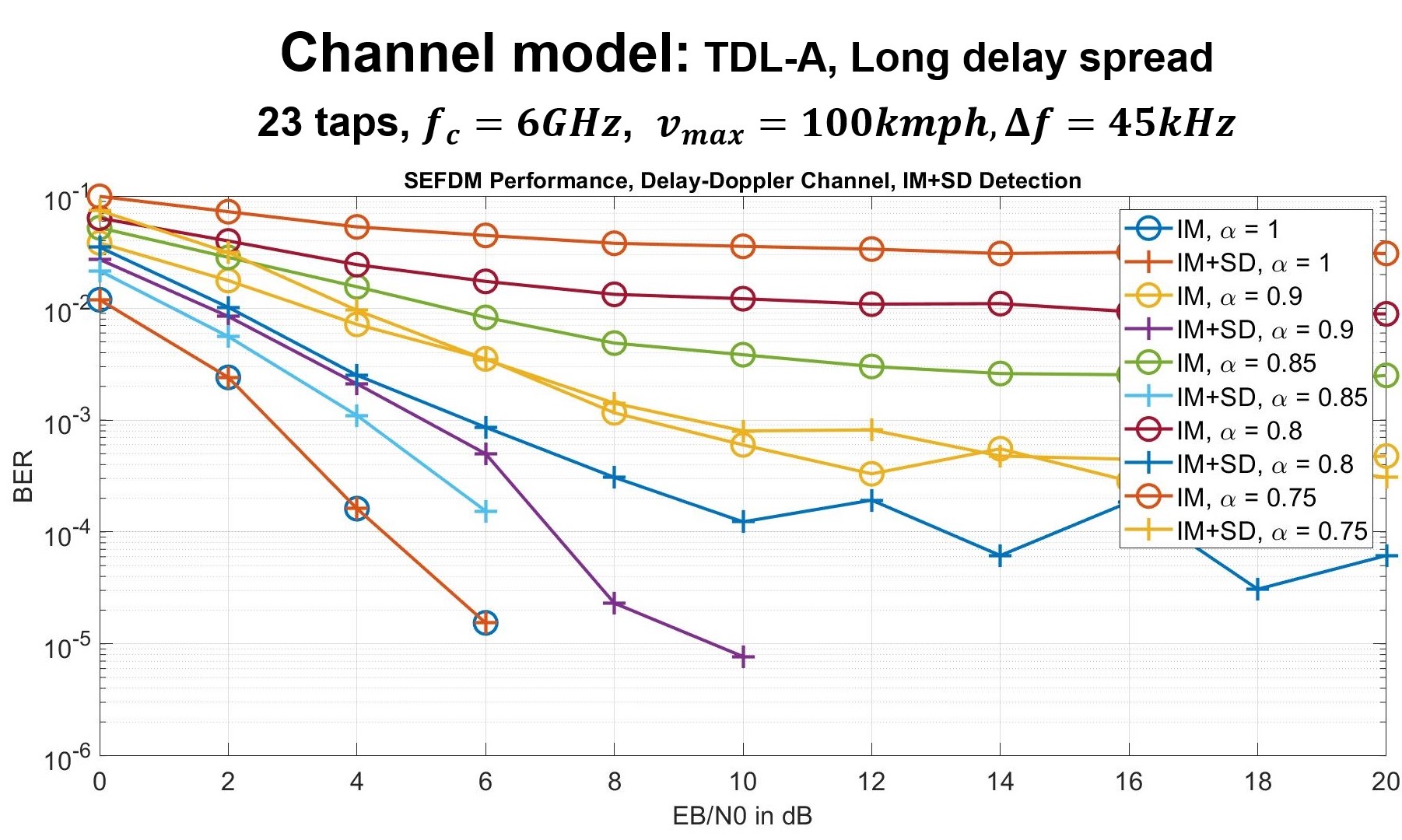}
		\subcaption{Hard decision making with
			$\lambda = 0.8$ and $n_{iter} = 20$.
		}
		\label{fig:SEFDM_DD_SD4}
		
	\end{subfigure}%
		\subcaption{
			Plot of {BER} in terms of {EB/N0} in signal detection
			{SEFDM}
			Using {IM} and {SD} algorithm in delay-Doppler channel.
		}
	\end{figure*}

In Fig. \ref{fig:SEFDM_DD_SD3}, employing the soft decision IM with a high delay spread and Doppler shift ($v_{max}=500 \, \text{km/h}$) proves inefficient in terms of BER. However, utilizing the FDCP method and employing SD with IM as the initial estimate yields a relatively low BER (up to $10^{-4}$ at compression factors $\alpha=0.9, 0.85$). It is noteworthy that initial error compensation is conducted using soft decision, with no application of hard decision. Fig. \ref{fig:SEFDM_DD_SD4} illustrates the efficacy of employing the hard decision IM with SD under the conditions of high delay spread and moderate Doppler shift ($v_{max}=100 \, \text{km/h}$). In such scenarios, hard decision in the OFDM performs well, achieving successful detection without SD. Applying hard decision to SEFDM in conjunction with the IM, and using it as the initial estimate for the SD, reduces BER to less than $10^{-3}$ and even less than $10^{-4}$ at compression factors $\alpha=0.9, 0.85, 0.8$.

\section{Conclusion}
\label{sect:con}
We initiated our study by studying a model of the SEFDM communication system and investigating the impact of using the MNL acceleration with soft and hard decision IM on the performance of SEFDM detection in the AWGN channel. Then, by adding FDCP to the SEFDM frame, we enhanced its resilience against Doppler shift in the channel. By using waveform shaping with RRC, we presented simulation results that were closer to reality, taking into consideration the standard TDL-A model, and reported BER detection figures using Regularized Sphere Decoding in various simulation scenarios. In this way, we demonstrated that it is possible to achieve acceptable performance in Doppler channels while maintaining the superiority of SEFDM over OFDM in terms of spectral efficiency. On the other hand, simulations revealed that using IM as an inverse system calculation algorithm can reduce the effect of delay spread even without using [time] CP. However, for very long delay spreads, as simulations show, this system cannot perform well even with the use of OFDM ($\alpha=1$).


\bibliographystyle{IEEEtran}
\bibliography{IEEEexample.bib}

\end{document}